\documentstyle[twocolumn,aps,epsfig,floats]{revtex}
\draft
\begin{document}
\twocolumn[\hsize\textwidth\columnwidth\hsize\csname @twocolumnfalse\endcsname

\title{Half-metallic antiferromagnets in thiospinels} 

\author {Min Sik Park and B. I. Min}

\address{Department of Physics, Pohang University of Science and
	Technology, Pohang 790-784, Korea}

\date{\today}
\maketitle

\begin{abstract}

We have theoretically designed the half-metallic (HM) antiferromagnets (AFMs)
in thiospinel systems, $\rm Mn(CrV)S_{4}$ 
and $\rm Fe_{0.5}Cu_{0.5}(V_{0.5}Ti_{1.5})S_{4}$,
based on the electronic structure studies 
in the local-spin-density approximation (LSDA). 
We have also explored electronic and magnetic properties
of parent spinel compounds of the above systems;
$\rm CuV_{2}S_{4}$ and $\rm CuTi_{2}S_{4}$ are found to be
HM ferromagnets in their cubic spinel structures, while
$\rm MnCr_{2}S_{4}$ is a ferrimagnetic insulator.
We have discussed the feasibility of material synthesis 
of HM-AFM thiospinel systems.

\end{abstract} 

\pacs{PACS numbers: 71.20.Be, 75.25.+Z, 75.50.Ee}
]

   
Since the first theoretical report of Heusler
half-metallic (HM) ferromagnet NiMnSb by
de Groot {\it et al.} \cite{degroot83},
much effort has been devoted to developing the HM 
magnetic materials, in which the conduction electrons at the 
Fermi level $\rm E_F$ are $100 \%$ spin-polarized \cite{irkhin}. 
Especially, the HM antiferromagnet (AFM) attracts great attention
because it is a non-magnetic metal but its conduction electrons are $100 \%$ 
spin-polarized.
It can be used as a probe of the spin-polarized scanning tunneling 
microscope without perturbing the spin-character of samples.
Further, the HM-AFM is expected to play a vital role   
in the future spintronic devices that utilize the spin polarization of 
the carriers.

The first HM-AFM, $\rm V_{7}MnFe_{8}Sb_{7}In$, which is
a derivative of the Heusler compound, was proposed by van Leuken 
and de Groot \cite{Groot}.
Anther possibility was suggested by Pickett \cite{Pickett} in 
the double perovskite system such as La$_2$VMnO$_6$.
In this case, V and Mn have antiferromagnetically aligned magnetic moments 
that exactly cancel each other.
To date, there has been no successful experimental realization of the HM-AFM.  

The thiospinel $\rm FeCr_2S_4$ in its metallic phase has
the HM ferrimagnetic state with nominal valence configurations of
Fe$^{2+}$ ($d^6$) and Cr$^{3+}$ ($d^3$) \cite{MSPark,Jskang,Kurmaev}.
The magnetic moments of Fe and Cr are $4 \mu_B$ and $-3 \mu_B$,
respectively, which produce the integer total magnetic moment of
$-2 \mu_B$ per formula unit.  
From this, one can expect that $\rm FeV_2S_4$ becomes a HM-AFM, 
since V has one less electron than Cr and so
the magnetic moment of two V$^{3+}$ ($d^2$: S=1) ions would cancel 
that of Fe$^{2+}$ ($d^6$: S=2), still possessing the HM property. 
Indeed the local-spin-density approximation 
(LSDA) band calculation yields the HM-AFM electronic structure of
$\rm FeV_2S_4$ \cite{Mspark01}.
Unfortunately, $\rm FeV_{2}S_{4}$ does not exist in the cubic spinel structure
but in the hexagonal NiAs structure ($Cr_{3}S_{4}$-type) 
with a complicated magnetic configuration \cite{Powell}.
So the above expectation does not work for $\rm FeV_2S_4$ in nature.

Motivated by the above expectation, we attempt to search for
the HM-AFM in other thiospinel compounds.
Most of the thiospinel compounds of $AB_{2}$S$_{4}$-type 
($A,B$: transition metals)
with cubic structure have a ferrimagnetic ground state. 
Usually, the magnetic moment of the $A$ ion in the tetrahedral site is 
antiferromagnetically  polarized with that of the $B$ ion
in the octahedral site. 
Under this circumstance, there are some pairs of $A$ and $B$ 
which give rise to the exactly cancelled magnetic moment in $AB_{2}$S$_{4}$. 
Moreover, some spinels have the HM electronic 
structures, as in $\rm FeCr_2S_4$. Then, by choosing proper pairs 
which satisfy these two conditions, 
one can devise the thiospinels with the HM-AFM nature.  
One possible pair is $\rm Mn (5\mu_{B})$ at A-site 
and mixed cations $\rm Cr (3\mu_{B})$-$\rm V (2\mu_{B})$ at B-site. 
Another possible pair is $\rm Fe_{0.5} (2.5\mu_{B})$-Cu$_{0.5} (0.0\mu_{B})$ 
at A-site and $\rm V_{0.5} (1\mu_{B})$-$\rm Ti_{1.5} (1.5\mu_{B})$ at B-site.
Even though a pair like $\rm Fe(4\mu_{B})$ and $\rm V(2\mu_{B})$ can be chosen,
they are not crystallized in the cubic spinel structure, as discussed above. 
For the same reason, the pairs, such as
$[\rm Co(3\mu_{B})$ and $\rm Ti(1\mu_{B})$-$\rm V(2\mu_{B})]$, and
$[\rm Ni(2\mu_{B})$ and $\rm Ti(1\mu_{B})]$, are not suitable for
a possible HM-AFM, because Ti and V form thiospinels
only with Cu \cite{Wold}.

In this work, we will propose two candidates of HM-AFM 
in thiospinels with cubic spinel structure: $\rm Mn[CrV]S_{4}$ and 
$\rm Fe_{0.5}Cu_{0.5}[V_{0.5}Ti_{1.5}]S_{4}$. 
We show that electronic structures of these thiospinels
have the HM-AFM nature, by using the LSDA on the basis of
the linearized muffin-tin orbital (LMTO) band method. 
The von Barth-Hedin form of the exchange-correlation potential is utilized.

\begin{figure}
\epsfig{file=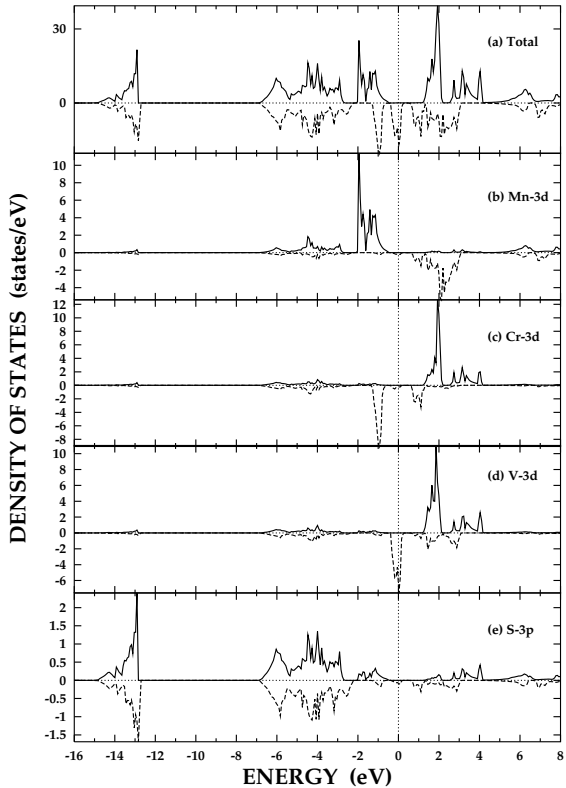,width=7.50cm}
\caption{Total and projected local density of states (PLDOS) 
         of $\rm Mn[CrV]S_{4}$.
}
\label{Mn}
\end{figure}


{\it $Mn[CrV]S_{4}$} $-$ 
We have performed the LSDA band calculation for
a hypothetical spinel $\rm Mn[CrV]S_{4}$.
We used two formula units in the primitive unit cell.
The primitive unit cell is formed by a cubic closed-packed (fcc) array
of S atoms, in which one eighth of the tetrahedral and one-half of the
octahedral interstitial sites are occupied by cations. S atoms have
four-fold coordination, formed by three octahedral cations and one
tetrahedral cation.
For the LMTO calculation, four empty spheres are considered in the
interstitial sites to enhance the packing ratio.

Figure~\ref{Mn} shows that $\rm Mn[CrV]S_{4}$ has 
the HM-AFM ground state at the  lattice constant $a=10.110$\AA 
~which is the experimental lattice constant of $\rm MnCr_{2}S_{4}$.  
Spins of Cr and V are ferromagnetically aligned, but
they are antiferromagnetically aligned to those of Mn.
The energy gap is evident near $\rm E_F$
in the spin-up density of states (DOS), and the DOS at $\rm E_F$ 
originates mainly from spin-down V-$3d$ ($\sim$ two $t_{2g}$ electrons)
states which will contribute to the metallic conductivity. 
Five spin-up Mn-$3d$ states and three spin-down Cr-$t_{2g}$ states
are fully occupied. 
Note that small spin-down states are induced in Mn and Cr near $\rm E_F$
by the hybridization with those of V.
The occupied DOS suggests that $\rm Mn[CrV]S_{4}$
has a nominal valence configuration of $\rm Mn^{2+}[CrV]^{3+}S_{4}^{2-}$,
which corresponds to the normal spinel structure.  
By counting the small magnetic moments of S and empty spheres (ES), 
the total magnetic moment becomes zero in the unit cell 
which is a nature of the HM-AFM (Table \ref{table1}).

\begin{figure}
\epsfig{file=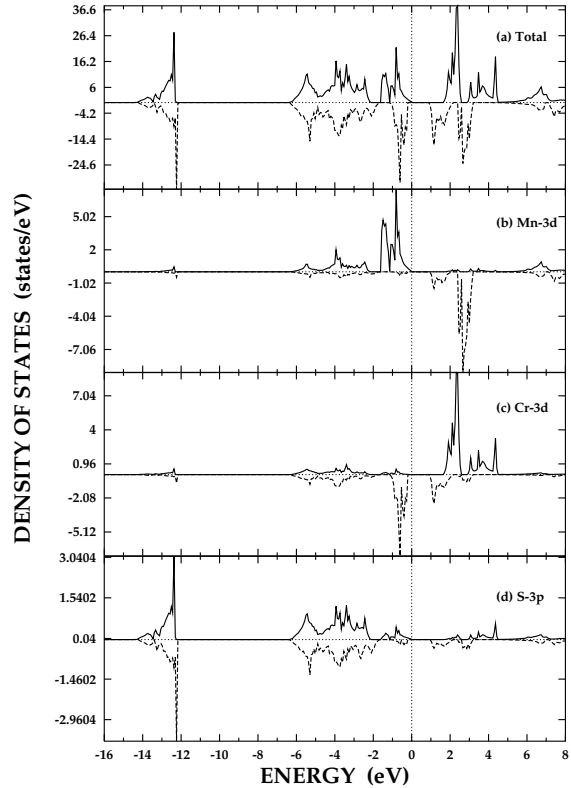,width=7.50cm}
\caption{Total and PLDOS of $\rm MnCr_{2}S_{4}$.
}
\label{MnC}
\end{figure}

$\rm MnCr_{2}S_{4}$, the end member of $\rm Mn[CrV]S_{4}$, is
known to be a ferrimagnetic insulator with $T_{C}=80$ K
\cite{Lotgering} and with the total magnetic moment $\sim 1\mu_{B}$
\cite{Tsurkan}. 
\begin{table}[b] \centering
\caption{\label{table1} Total and local spin magnetic moments in $\mu_{B}$
        of $\rm Mn[CrV]S_{4}$ (MCVS) and $\rm MnCr_{2}S_{4}$ (MCS).
   ES denotes the empty sphere.}
  \begin{tabular}{ccccccc}
       &Total & Mn & Cr & V & S & ES \\ \hline
   MCVS&0.00 & $-$4.55 & 3.17 & 1.99 & $-$0.11 & $-$0.09 \\
   MCS &1.00 & $-$4.56 & 3.13 &      & $-$0.13 & $-$0.09 \\
  \end{tabular}
\end{table}
Neutron diffraction experiment at 4.2 K provided
that Cr and Mn have local magnetic moments of $3\mu_{B}$ and
$4.7\mu_{B}$, respectively \cite{Menyuk}.  
Our LSDA band calculation yields
the insulating ground state (Fig.~\ref{MnC}) with local magnetic moments of
$3.13\mu_{B}$ and $4.56\mu_{B}$ for $\rm Cr$ and $\rm Mn$, respectively,
which agrees well with  experimental results (Table \ref{table1}).
The insulating property comes from 
the $\rm Mn$ $(e_{g}^{2}t_{2g}^{3})$ exchange split spin polarized 
energy gap and the $\rm Cr$ $(t_{2g}^{3})$ crystal field split
energy gap. The ferrimagnetism in $\rm MnCr_{2}S_{4}$ can be understood
in terms of the ordinary kinetic superexchange interaction.
It is thus evident that the HM-AFM $\rm Mn[CrV]S_{4}$ results from the
insertion of V-$3d$ band in the gap region of $\rm MnCr_{2}S_{4}$
with the magnetic moment parallel to that of Cr.

Unfortunately, the other end member $\rm MnV_{2}S_{4}$  
exists in the $Cr_{3}S_{4}$-type structure \cite{Nogues}.
It has been reported that the spinel phase of $\rm Mn[Cr_{2-x}V_{x}]S_{4}$
is formed successfully up to $x=0.6$ with decreasing magnetic moment from 
$\sim 1\mu_{B}$ at $x=0.0$ to $\sim 0.4\mu_{B}$ at $x=0.6$ \cite{Goldstein}.
From this, one can extrapolate zero magnetic moment
at $x=1.0$ in accordance with the present LSDA band result.
Only when the spinel structure
with substituting $\rm V$ for $\rm Cr$ in $\rm MnCr_{2}S_{4}$ is retained, 
the HM-AFM $\rm Mn[CrV]S_{4}$ could be synthesized.
The above experimental result may indicate that the bulk spinel phase 
of $\rm Mn[Cr_{2-x}V_{x}]S_{4}$ would be unstable over $x=0.6$.
Then the possible attempt would be an artificial fabrication of 
$\rm Mn[CrV]S_{4}$ film on appropriate substrates, 
which is desirable to be tested experimentally.

\begin{figure}
\epsfig{file=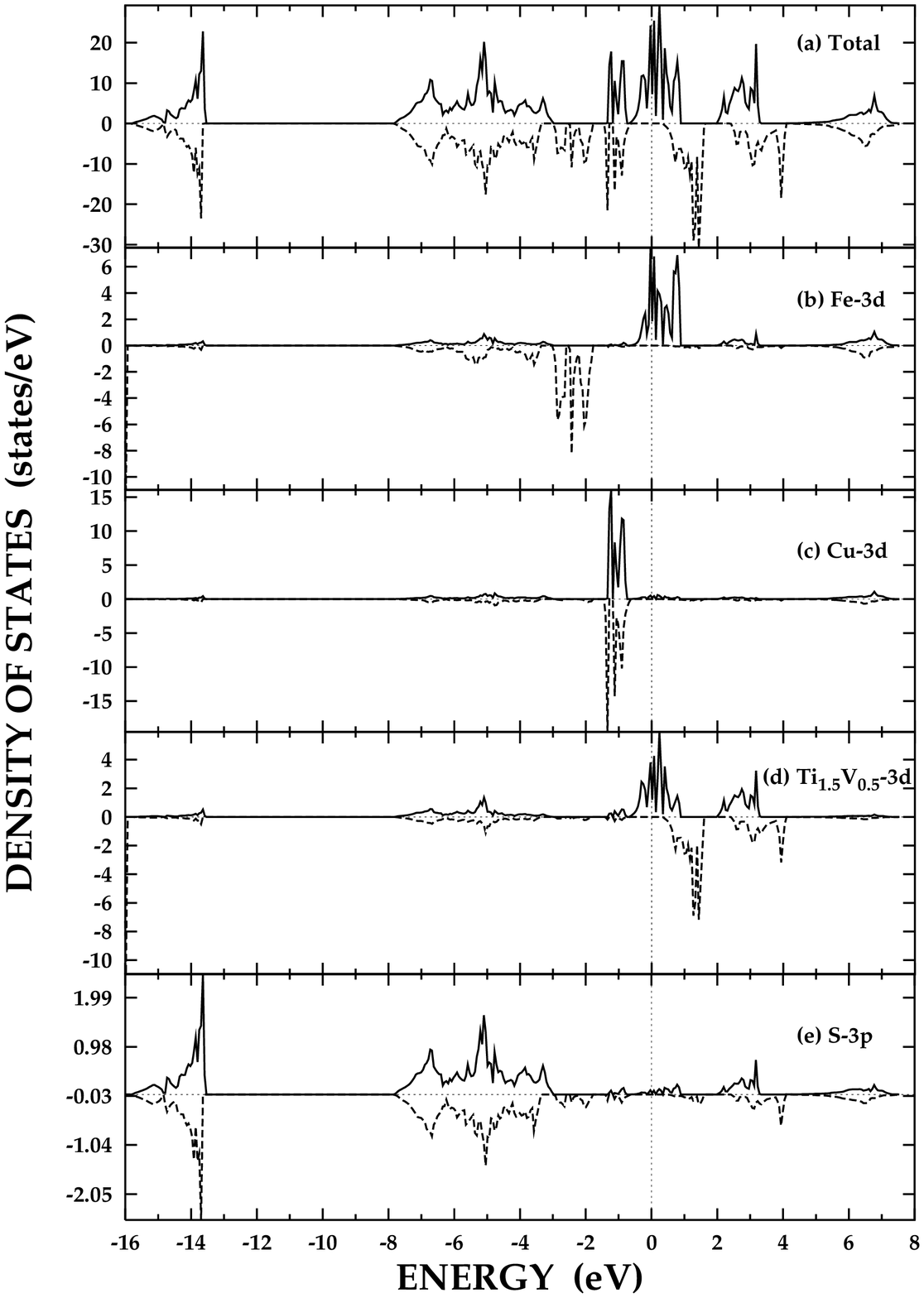,width=7.50cm}
\caption{Total and PLDOS 
	of $\rm Fe_{0.5}Cu_{0.5}[V_{0.5}Ti_{1.5}]S_{4}$.
 }
\label{Fe}
\end{figure}

{\it $Fe_{0.5}Cu_{0.5}[V_{0.5}Ti_{1.5}]S_{4}$} $-$
We have mentioned that $\rm FeV_{2}S_{4}$, if it exists in the cubic spinel
form, would be a strong candidate for the HM-AFM.
We have also seen that V and Ti form thiospinels only with Cu:
$\rm CuV_{2}S_{4}$ and $\rm CuTi_{2}S_{4}$ \cite{Wold}.
On the basis of this, we have explored the combination of Fe, Cu, V, 
and Ti to make a thiospinel possessing both the HM and the AFM nature:
$\rm Fe_{0.5}Cu_{0.5}[V_{0.5}Ti_{1.5}]S_{4}$ (FCVTS).
We have performed the LSDA electronic structure calculation for FCVTS
by replacing 
$\rm V_{0.5}Ti_{1.5}$ with a virtual atom having atomic number 22.25. 
The lattice constant of $a=9.8865$\AA ~is used which is interpolated from
end members, $\rm CuV_{2}S_{4}$ and $\rm CuTi_{2}S_{4}$.
As shown in Fig. \ref{Fe}, FCVTS really has the HM-AFM
electronic structure.
Ten Cu $3d$-states and five spin-down Fe $3d$-states are fully occupied.
The DOS at $\rm E_F$ originates from  spin-up Fe $3d$ 
($\sim$ $e_{g}^1$) and $\rm V_{0.5}Ti_{1.5}$ $3d$
($\sim$ $t_{2g}^2$) states.
\begin{table}[b] \centering
\caption{\label{table2}  Total and local spin magnetic moments in $\mu_{B}$ 
	of $\rm Fe_{0.5}Cu_{0.5}[V_{0.5}Ti_{1.5}]S_{4}$ (FCVTS),
         cubic spinel $\rm CuV_{2}S_{4}$ (CVS), and
         $\rm CuTi_{2}S_{4}$ (CTS).
}
  \begin{tabular}{ccccccccc}
   &Total & Fe & Cu & $\rm V_{0.5}Ti_{1.5}$ & V & Ti & S & ES \\ \hline
   FCVTS&0.00 & $-$3.57 & $-$0.13 & 1.08 &  &  & $-$0.07 & 0.03 \\
   CVS&3.00 & & $-$0.28 & & 1.89 & & 0.12 & 0.00 \\
   CTS&1.00 & & $-$0.17 & & & 0.66 & 0.04 & 0.00 \\
  \end{tabular}
\end{table}
The calculated magnetic moments in Table.~\ref{table2}
reflect that the valence configuration of FCVTS corresponds to
$\rm Fe_{0.5}^{2+}Cu_{0.5}^{1+}[V_{0.5}Ti_{1.5}]^{3.25+}S_{4}^{2-}$
with zero total magnetic moment.
The magnetic mechanism in FCVTS can be understood in a similar way
to the case in $\rm Fe_{0.5}Cu_{0.5}Cr_{2}S_{4}$ \cite{MSPark}

\begin{figure}
\epsfig{file=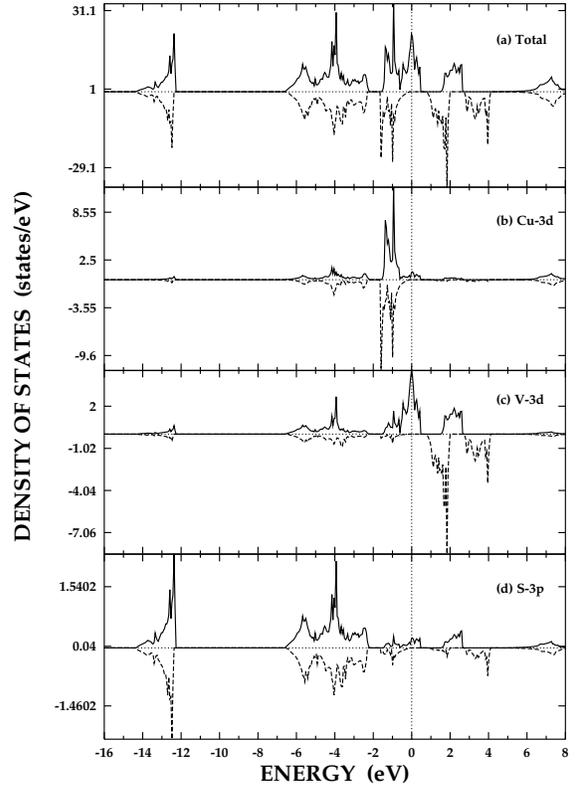,width=7.50cm}
\caption{Total and PLDOS of $\rm CuV_{2}S_{4}$.
}
\label{CuV}
\end{figure}

Now, let us discuss the possibility of synthesizing FCVTS.
We first consider parent thiospinels, $\rm CuV_{2}S_{4}$ and 
$\rm CuTi_{2}S_{4}$, both of which have metallic nature \cite{Bouchard}.
The LSDA band calculation shows that the $\rm CuV_{2}S_{4}$ has a 
HM ground state in its cubic spinel structure
(Fig.~\ref{CuV}).  
But, this result does not coincide with experimental results for
$\rm CuV_{2}S_{4}$ having no spontaneous magnetic property.
In fact, $\rm CuV_{2}S_{4}$ has a CDW-type phase transition below 90 K.
Accordingly, the cubic unit cell is transformed to a tetragonal 
structure which is induced by the Jahn-Teller local small distortion 
at the V site \cite{Yoshikawa,Fleming,Mahy,Hagino,Ohno}.
Our result suggests that if $\rm CuV_{2}S_{4}$ exists in the cubic spinel 
structure, it will have a HM ferromagnetic ground state.
The oxidation state of Cu atoms of $\rm CuV_{2}S_{4}$ is 
closer to $\rm Cu^{+}$ \cite{Lu}. The bands near $\rm E_F$
are predominantly of $\rm V$ $3d$ character, so that the metallic 
nature of $\rm CuV_{2}S_{4}$ comes mainly
from the $\rm V$ atom\cite{Lu}.

The CDW-type phase transition can be suppressed by the random occupation 
of the octahedral sites in the spinel structure.
Indeed, $\rm Cu[V_{1.9}Ti_{0.1}]S_{4}$ showed no evidence of a phase 
transition due to random occupation of the octahedral sites
by the V-Ti mixture\cite{Disalvo}.
$\rm CuV_{2}S_{4}$ has a large susceptibility due to 
considerable exchange enhancement \cite{Disalvo}. 
The ferromagnetism in the Cr-mixed $\rm CuV_{2}S_{4}$ compound 
$\rm CuV_{2-x}Cr_{x}S_{4}$ at low Cr concentrations is considered 
to come from the large 
susceptibility of $\rm CuV_{2}S_{4}$\cite{Disalvo}. Therefore,
one can also expect that, if some of Cu atoms are replaced by Fe atoms,
$\rm (FeCu)V_{2}S_{4}$ would become a ferromagnet.

\begin{figure}
\epsfig{file=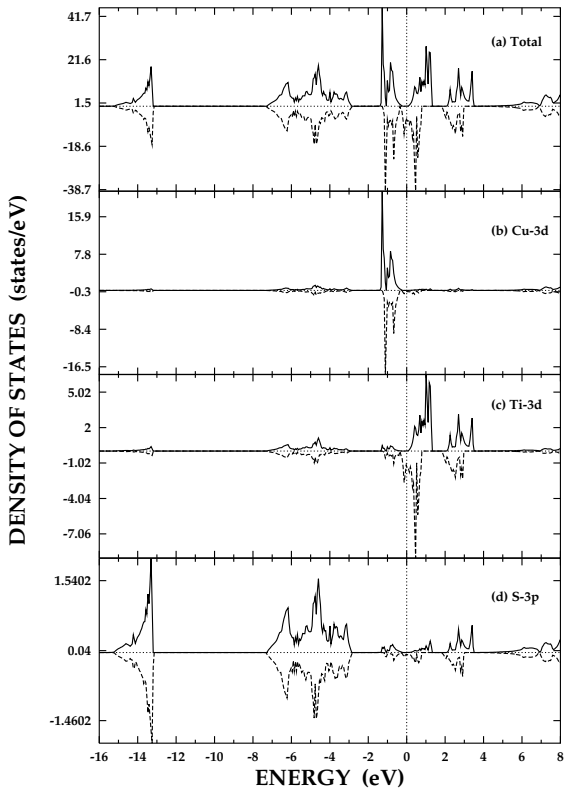,width=7.50cm}
\caption{Total and PLDOS of $\rm CuTi_{2}S_{4}$.
}
\label{CuTi}
\end{figure}

$\rm CuTi_{2}S_{4}$ shows a metallic behavior \cite{Matsumoto}
and the weak magnetism \cite{Koyama}.
Our LSDA band calculation also indicates that both the metallic nature
and the weak magnetism of $\rm CuTi_{2}S_{4}$ result from 
$3d$-bands of $\rm Ti$ ions (Table.~\ref{table2}).
As shown in Fig.~\ref{CuTi}, the $\rm CuTi_{2}S_{4}$ also has a HM 
ferromagnetic ground state.
The electronic configuration of $\rm Cu$ at the A-site is considered 
to be close to $e_{g}^{4}t_{2g}^{6}$ (S=0).
Combining the above features, one can make a metallic 
$\rm Cu[V_{0.5}Ti_{1.5}]S_{4}$ having a large susceptibility.
Then by mixing $\rm Fe$ ions with $\rm Cu[V_{0.5}Ti_{1.5}]S_{4}$
to have exact cancellation of magnetic moments,
one can possibly make HM-AFM FCVTS.


In conclusion, we have predicted that $\rm Mn[CrV]S_{4}$ and FCVTS
are candidates for the HM-AFM in thiospinels, based on the LMTO band 
results in the LSDA.
$\rm Mn[CrV]S_{4}$ and FCVTS satisfy two criteria for the HM-AFM: 
the antiferromagnetic couplings between magnetic moments and the HM property.
The possibility of synthesizing HM-AFMs is discussed by considering 
electronic, transport, and magnetic properties of parent thiospinel compounds. 
$\rm MnCr_{2}S_{4}$ found to be a ferrimagnetic insulator in agreement 
with experiment, while $\rm CuV_{2}S_{4}$ and $\rm CuTi_{2}S_{4}$ are 
found to be HM ferromagnets in their cubic spinel structures.

Acknowledgments $-$ 
This work was supported by the KOSEF through the eSSC at POSTECH
and in part by the BK21 Project.

\end{document}